\documentclass[
    ,final            
  ]
  {aipproc}

\layoutstyle{8x11double}

\def\int{{\it INTEGRAL}}
\def\rxs{1RXS J170849.0--400910}
\def\zerosei{SGR 1806--20}

\begin{document}

\title{Hard X-ray variability of Magnetar's Tails observed with INTEGRAL}

\classification{97.60.Gb; 98.70.Rz}
\keywords      {pulsars; magnetars; hard X-rays}

\author{D. G\"{o}tz}{
  address={CEA Saclay, DSM/Dapnia/Service d'Astrophysique, F-91191, Gif sur Yvette, France}
}

\author{S. Mereghetti}{
  address={INAF--Istituto di Astrofisica Spaziale e Fisica Cosmica Milano, via Bassini 15, I-20133 Milano, Italy}
}

\author{K. Hurley}{
  address={University of California at Berkeley, Space Sciences Laboratory, Berkeley CA 94720-7450, USA}
}

\author{P. Esposito}{
  address={INAF--Istituto di Astrofisica Spaziale e Fisica Cosmica Milano, via Bassini 15, I-20133 Milano, Italy}
  ,altaddress={University of Pavia, Department of Nuclear and Theoretical Physics and INFN-Pavia, Pavia, Italy} 
}

\author{E. V. Gotthelf}{
  address={Columbia Astrophysics Laboratory, Columbia University, 550 West 120th Street, New York, NY 10027, USA}
}

\author{G. L. Israel}{
  address={INAF--Osservatorio Astronomico di Roma, Via Frascati 33, I-00040 Monteporzio Catone (Roma), Italy}
}

\author{N. Rea}{
  address={University of Amsterdam, Astronomical Institute "Anton Pannekoek", 1098~SJ, Amsterdam, The Netherlands}
}

\author{A. Tiengo}{
  address={INAF--Istituto di Astrofisica Spaziale e Fisica Cosmica Milano, via Bassini 15, I-20133 Milano, Italy}
}

\author{R. Turolla}{
  address={University of Padua, Department of Physics, via Marzolo 8, 35131 Padova, Italy}
  ,altaddress={Mullard Space Science Laboratory, University College London, Holmbury St. Mary, Droking Surrey, UK} 
}

\author{S. Zane}{
  address={Mullard Space Science Laboratory, University College London, Holmbury St. Mary, Droking Surrey, UK}
}

\begin{abstract}
 Magnetar's persistent emission above 10 keV was recently discovered thanks to the imaging
capabilities of the IBIS coded mask telescope on board the \int~ satellite. The only two sources
that show some degree of long term variability are \zerosei~ and \rxs. We find some
indications that variability of these hard tails could be the driver of the spectral variability
measured in these sources below 10 keV.

In addition we report for the first time  the detection at 2.8 $\sigma$ level of pulsations in the hard X-ray 
tail of \zerosei.
\end{abstract}

\maketitle

\section{Introduction}

 Anomalous X-ray Pulsars (AXPs) and Soft Gamma-Ray Repeaters (SGRs) are two small
classes of sources that are believed to be magnetar candidates, namely isolated neutron stars
with magnetic fields larger than the quantum critical value B$_{QED}\sim$4.4$\times$10$^{13}$G.
For a review of this class of objects see \cite{woodsrew}.

They share some common properties like a long spin period in the range of $P$=2--12 s, a large period
derivative of $\dot P$=10$^{-13}$--10$^{-10}$ s s$^{-1}$, and a typical X-ray luminosity 
of $L_{X}\sim 10^{34}$--10$^{36}$ erg s$^{-1}$. This luminosity is well above the rotational energy 
losses, and is believed to be powered by the decay of the huge magnetic field.

Since the spectra below $\sim$10 keV are rather soft, the
first \int~ detections above 20 keV of very hard high-energy tails
associated with these objects came as a surprise
\citep{kuiper,denhartog,rev,mere05,molkov,dg06}. 
The spectra flatten ($\Gamma\sim$ 1, where $\Gamma$ is the photon index)
above 20 keV and the pulsed fraction of some of them reaches up to 100\% \citep{kuiper}. 
The discovery of these hard tails provides new constraints on the emission models for these objects
since their luminosities might well be dominated by hard, rather than soft, X-rays.

In this paper we will focus on the timing properties and long term variations of two magnetar candidates, \rxs~ and \zerosei, as measured by IBIS/ISGRI \cite{ibis,isgri} on board the \int~ satellite \cite{integral}.

\section{\rxs}
\rxs~ was discovered during the {\it ROSAT} all sky survey. The measure of
its period \citep{sugizaki}, period derivative \citep{kaspi1708} and
general X-ray properties \citep{giallo}, made it an AXP member.
Interesting results have been reported by \cite{nanda}, who claimed a
correlation between the soft X-ray flux and spectral hardness, with a marginal evidence of
the highest and hardest spectra being correlated with the AXP glitching activity
\citep{simone03,ka00,luanasim}.
This correlation has recently been confirmed by \cite{campana}, \cite{nanda07},
and \cite{luanasim} using further {\it Chandra}, {\it Swift}/XRT and {\it RXTE} data,
but always focusing on a limited energy interval (i.e. below $\sim 10$~keV).

Recently (fall 2006 and spring 2007) we performed a multi-wavelength observation
campaign aimed at monitoring \rxs~ with \int~ and {\it Swift}
\cite{swift}. These new data, together with a re-analysis of all
publicly available \int, and soft X-ray observations,
allowed us to investigate the timing and spectral properties of the
source over a broader energy range. We discovered a long term correlation between
the soft and hard X-ray emission. 

\begin{figure}[ht!]
  \includegraphics[width=7.cm]{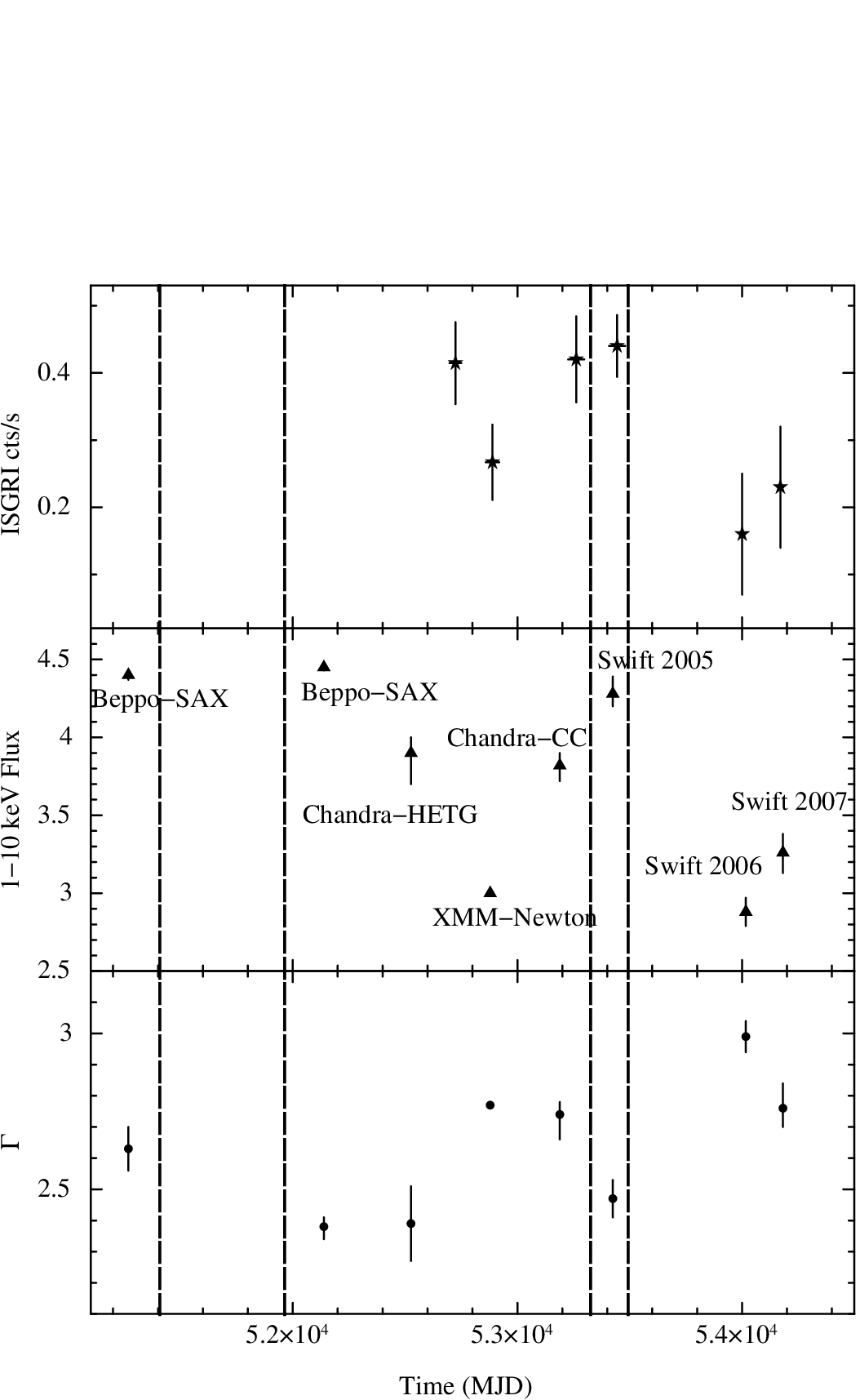}
  \caption{Upper panel: hard X-ray fluxes derived with \int/IBIS (20--70 keV). Middle Panel: absorbed 1-10 keV fluxes in units of 10$^{-11}$ erg cm$^{-2}$ s$^{-1}$ derived from recent observations of X-ray imaging telescopes as a function of time. Lower Panel: photon indices measured in the 1-10 keV energy band. Vertical lines mark the times of four observed glitches. Errors are reported at 1$\sigma$ c.l. From \cite{dg07}.}
\label{fig:corr}
\end{figure}

The spectral parameters (fluxes and $\Gamma$) plotted in Fig. \ref{fig:corr} have been derived
by fitting all the datasets simultaneously in 1--10~keV energy range (except for
{\em Chandra} data which were limited to 8\,keV) by using an absorbed black body plus power
law model.
While the parameters of the power law have been left free to vary, the
absorption column density, and the black body temperature have been forced to be the same
for all instruments. For the details of our analysis see \cite{dg07}.
We found a good fit ($\chi^2$/d.o.f. =  1188/1264 = 0.94),
and the derived values were $N_{H}$=1.36(1)$\times$10$^{22}$ cm$^{-2}$ 
, and $kT= 0.44(1)$~keV.
Our new XRT observation campaign (last two points in Fig.~\ref{fig:corr})
shows that the source entered a new low/soft state (similar to the one measured with {\it
XMM-Newton} in 2003), with a flux a factor $\sim$1.5 lower than
the one measured with XRT in 2005. By adding these new points to the long
term variability study of the source, we confirm the flux-hardness
correlation proposed by \cite{nanda}. 

\vspace{-2cm}


In addition, our new hard-X data show that the long term variation in flux
is correlated over more than two orders of magnitude in energy. In
fact, IBIS observations taken quasi-simultaneously with
the last two XRT ones indicate that
the source is hardly detected above 20 keV, while the
hard X-ray count rates measured before
followed well the variations measured in the soft X-ray range.
Unfortunately, due to the faintness of the source we could not statistically prove
spectral changes at high energies, by comparing different IBIS observations.

We also tested if the spectral variations at low energies
could be induced just by the variation of the high energy power law.
Indeed, by fixing the spectral parameters derived from the {\it
XMM-Newton} observations, one can fit the broad band {\it XMM/INTEGRAL}
(2003) and {\it Swift/INTEGRAL} observations (2005) by simply changing the
slope and normalization of the high energy power law. Unfortunately, due
to the uncertainties in the inter-calibration, and to the low statistic of the
{\it Swift} data, we can not draw a firm conclusion on this point.

Our timing analysis consisted in looking for pulsations in the IBIS data using the PCA ephemeris
derived by \cite{luanasim}. We computed the pulse profile of the source, as
a function of energy (see Fig. 1 in \cite{dg07}). We found that the peak
which is predominant in the soft band disappears and a secondary peak
grows with energy becoming the most prominent one above $\sim8$~keV. The
pulsed fraction cannot be easily derived from the PCA (a non-imaging
instrument) data due to the uncertainties in the background estimation. Besides,
the IBIS data are not easy to handle, because of large background
variations due the to fact that the source is at different off-axis angles during the different pointings.
Nevertheless, by averaging the background over
the entire IBIS observations, we could measure a pulsed fraction
by dividing the difference between the maximum and the minimum of the folded
light curves by the count rates derived from imaging. The resulting values
are: $\sim$25\% (in the 20--60 keV band), $\sim$60\% (60-200
keV), and $\sim$40\% (20--200 keV). Consistent results came from our analysis
of SAX/PDS archival data.

\section{\zerosei}
\zerosei~ was  discovered by the Interplanetary Network (IPN) in 1979 \cite{1806disc}.
\citet{kouve98} discovered pulsations ($P$=7.48 s) in the quiescent X-ray counterpart, which
was rapidly spinning down ($\dot P$=2.8$\times$10$^{-11}$ s s$^{-1}$). If this spin down is interpreted
as braking by a magnetic dipole field, its strength is $B\sim$10$^{15}$ G. The source activity is variable, 
alternating between quiet and very active periods.

\begin{figure}[h]
  \includegraphics[width=5.cm,angle=-90]{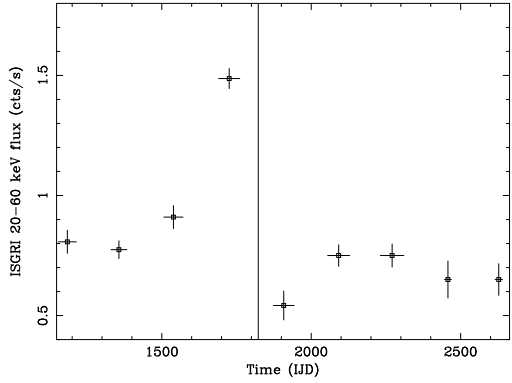}
  \caption{Hard X-ray persistent flux of \zerosei, as measured by IBIS/ISGRI in the 20-60 keV energy range. The
vertical line corresponds to the giant flare of December 27$^{th}$ 2004. The time is expressed in INTEGRAL Julian Day (=MJD-51544)}
\label{fig:monitoring}
\end{figure}

\zerosei~ is observed frequently by IBIS, thanks to its large field of view 
(29$^{\circ}\times$29$^{\circ}$) coupled to the fact that the source lies close to the Galactic
Center, one of \int's favorite targets. Thanks to IBIS high sensitivity we were able to discover and monitor
its high energy flux since 2003, showing that it is correlated with the source's bursting activity,
see e. g. \cite{dg07a}. We recently analyzed \int~ Key Programme data, taken
in fall 2006 and spring 2007. We confirm that the persistent flux has reached a level comparable with the one prior
to the giant flare, see Fig. \ref{fig:monitoring}. Despite the flux variations all the spectra have the same slope
($\Gamma\sim$2) in the 20-200 keV energy range, indicating that also in this case the hardening seen below
10 keV \cite{xmm} could be due just to the flux variation of the underlying hard component.
 
\begin{figure}[h]
  \includegraphics[width=5.cm,angle=-90]{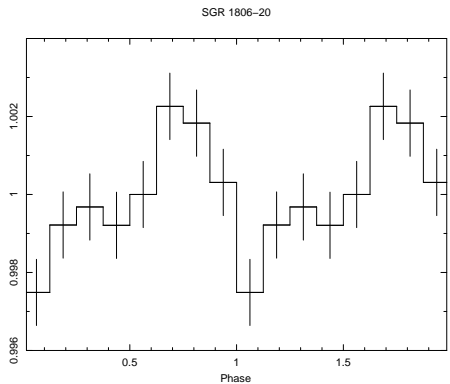}
  \caption{\zerosei~ folded light curve as measured with IBIS/ISGRI in the 20-200 keV energy band.}
\label{fig:puls}
\end{figure}

Making use of the ephemeris recently published by \cite{woods07} we folded \int~ data of \zerosei, in order to
look for pulsations at high energy. For the time being we restricted ourselves to the time period preceding the
giant flare where the flux is highest. Using $Z^{2}$ statistics we detected the pulsations at 2.8$\sigma$ level. A folded
light curve is shown in Fig. \ref{fig:puls}. Due to the complex background variations induced by the multitude
of variable sources in the Galactic Center region, the determination of the pulsed fraction is particularly delicate.
This work is on going, and the preliminary results indicate that the  pulsed fraction of \zerosei~ is smaller
than the one measured in the AXPs, similarly to what measured at soft X-rays.

\begin{theacknowledgments}
 D.G. acknowledges financial support from the French Space Agency (CNES).
\end{theacknowledgments}


\bibliographystyle{aipproc}   



\end{document}